\documentclass[twocolumn]{svjour3}        

\smartqed  

\usepackage{graphicx,amsmath}

\begin{document}


\title{Still Few-Nucleon Systems - After So Many Years
\thanks{Presented at the 21st European Conference on Few-Body Problems in Physics, 
Salamanca, Spain, 30 August - 3 September 2010}
}

\author{Peter U. Sauer}

\institute{ Peter U. Sauer  \at
Leibniz Universit\"{a}t Hannover, 
D-30167 Hannover, Germany \\
              \email{sauer@itp.uni-hannover.de} }

\date{Received: date / Accepted: date}

\maketitle


\begin{abstract}
Three- and four-nucleon reactions are discussed. The focus is on the notion of nuclear potentials, on conceptual and technical issues of calculations and  on unresolved problems between existing theoretical predictions and experimental data. The special focus is on some historic aspects of the evolution of the field. The views presented have a strong personal bias.
\end{abstract}

\section{Preliminaries \label{sec:1}}  

\begin{figure}[!]
\begin{center}
\includegraphics[scale=0.50]{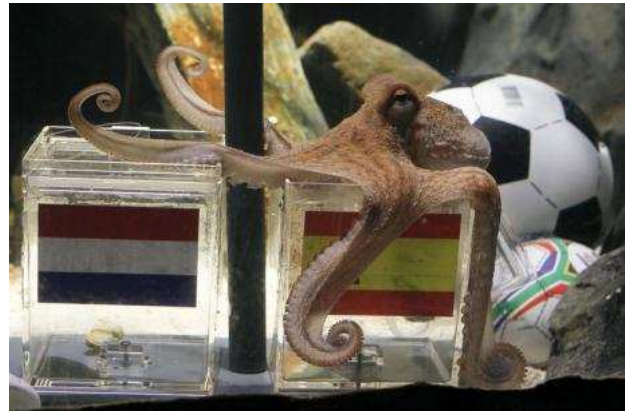} 
\end{center}
\caption{\label{fig:PulvoPaul} 
Octopus Paul predicting Spain to become the 2010 World Champion in Football \cite{paul:10}. Paul lived in the Aquarium Oberhausen, Germany; he passed away on 26.10.2010 with the age of almost three, after a full octopus life. As all of his species, he had nine brains and three hearts. He was a real prophet with the specialization in international football. He had fans everywhere in the world, but mostly in Spain. He was an honorary citizen of the town 
Carballi\~{n}o, Galicia, Spain.}
\end{figure}

This is neither a summary nor a scientific talk. Nor is it an attempt to explore the future of few-nucleon physics: In this respect, another German, octopus Paul, would be more reliable, and he delighted our Spanish friends with his precise  prediction of Fig.~\ref{fig:PulvoPaul}. I shall use my own research results as basis for this talk, but I shall tell you anecdotes, hopefully stories of more general interest, with a historic and very personal touch: {\it How did I survive for so long in our research field and survive all those conferences?}
 
I often call my more serious talks  {\it Few-Nucleon Systems - Test of Nuclear Dynamics}, the theme of these conferences from the beginning. When attending my first conference of this type, the 1972 UCLA Conference, Keith A. Brueckner \cite{brueckner:72} gave a plenary talk, and he emphasized this aspect. Brueckner was my hero in those days. I had just learnt nuclear many-body theory, had learnt the subtleties of Brueckner's theory, dealing with a strongly repulsive core in the two-nucleon (2N) potential. I had done nuclear matter, closed-shell nuclei and the shell model and was deeply frustrated by the limitations of my results. What did the disagreements between data and theoretical predictions tell me? Were they due to our approximate solution of the many-body problem or due to deficiencies of the interaction about which I wanted to learn? The computation of three-nucleon clusters in the nuclear medium was the stumbling block for theory at that time, and I had learnt about Faddeev, not in the context of scattering theory, but from Bethe's reformulation \cite{bethe:67} of those three-nucleon processes with the help of connectedness, in contrast to the original formulation in powers of the reaction matrix; that reformulation are the Bethe-Faddeev equations. At that conference I learnt, that three-nucleon clusters in the medium must be much more difficult than the 3N bound state, its many-body aspect being well controlled; a technically clean result on the 3N bound state should provide a reliable and definitive physics evaluation of the assumed nuclear force. I saw the promised land and decided to take a leave of absence from nuclear structure for 2 to 3 years, and I wanted to return then to nuclear structure with improved technical equipment. I never returned. But the idea, that techniques of few-nucleon systems could fertilize other fields, was already there. Indeed, few-body physics has turned quite diverse since those early days, and this conference is witness for that fact.
\begin{figure}[!]
\begin{center}
\includegraphics[scale=0.80]{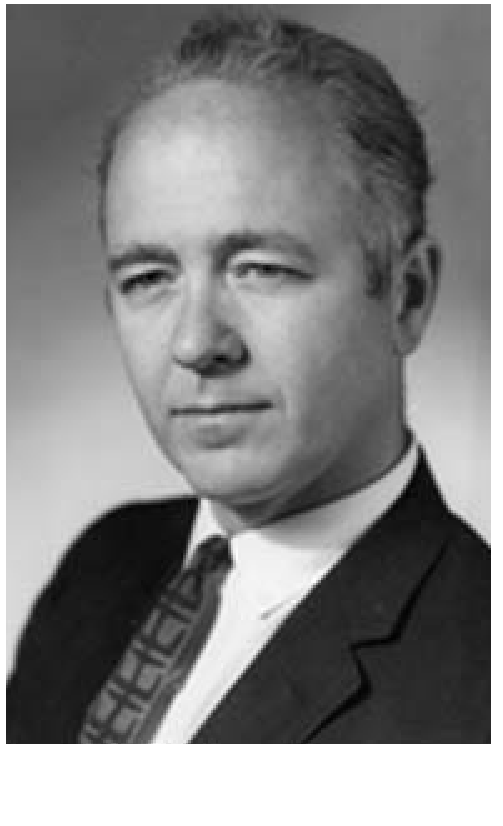}
\end{center}
\caption{\label{fig:Brueckner} 
Keith A. Brueckner, University of California, San Diego \cite{brueckner:10}.}
\end{figure} 

I tell stories with contact to my own research on 3N and 4N bound states and reactions.  Sect.2 describes  the nuclear dynamics chosen by me for the test in few-nucleon systems. Sect.3 discusses the Coulomb problem in scattering theory. Sect.4 gives results and Sect.5 conclusions. Sect.6 is an afterthought on the evolution of European few-body physics.


\section{Nuclear Dynamics \label{sec:2}} 

The November revolution of physics was in 1974: Charmonium was discovered, and soon after quantum chromodynamics (QCD) was identified as the underlying truth for nuclear dynamics. But it is already in the bible: Knowledge did not make life easier for mankind, it resulted in the expulsion from paradise. QCD also did not make life easier for nuclear theory, though it had an immediate impact on few-nucleon physics: In many existing problems the beneficial working of 6-quark bags was discovered by courageous colleagues; one of many examples is Ref. \cite{pirvar:81}. However, a realistic description of few-nucleon systems required the sober return to nucleons and potentials. In the beginning, potential forms were motivated by quark models, e.g.,  in Ref. \cite{valcarce:94}. In these days, the potentials of chiral effective field theory (EFT) \cite{kolck:02,epel:02}, taking into account constraints arising from QCD, are most popular \cite{machleidt:10}.

\begin{figure}[!]
\begin{center}
\includegraphics[scale=0.33]{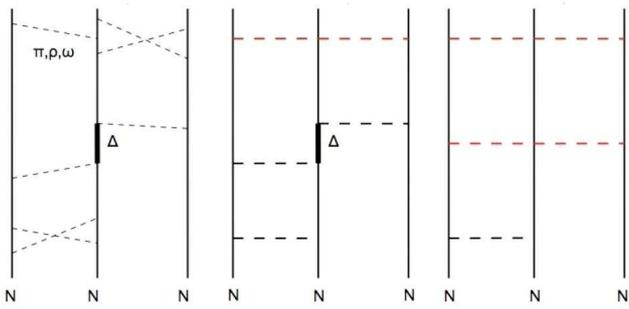}
\end{center}
\caption{\label{fig:Sauer-fig1} 
2N and 3N potentials. 
When reading the interacting 3N system from left to right, the figure shows how non-nucleonic degrees of freedom are frozen into instantaneous potentials by decreasing resolution. When reading from right to left, the figure shows how formerly unresolved non-nucleonic degrees of freedom are revived by increasing resolution; the middle part shows how one irreducible part of the 3N potential gets resolved into the successive action of a two-baryon coupled-channel potential. }
\end{figure}

The game is: Choose the nuclear dynamics and test your choice in few-nucleon systems. For the choice, there were always two aspects important: the {\it resolution} of the physics to be described and the {\it consistency} between 2N and many-N forces. Potentials are descriptive tools, babies of theoreticians, not created by nature or by God and therefore not experimentally measurable objects. I show you an old picture of mine in Fig.~\ref{fig:Sauer-fig1}, using  old-fashioned field theory with baryons and mesons for the transparency of argument: With decreased resolution the time-delayed meson exchanges get frozen into instantaneous potentials; by further decrease of resolution even nucleonic resonances cannot be distinguished. But there are different levels of resolution; we chose \cite{sauer:86} to keep the ${\Delta}$ isobar, derived from the nucleonic 3-quark bag just by a spin-isospin flip, alive and thereby resolved the 3N force partially into a coupled-channel interaction. I think, in EFT the resolution issue for the dynamics is similar: One may work in the leading order (LO) description or in sub-LOÔs, in a pionless or a pion\-ful description, and the forms of the potentials and the balance between 2N and many-N forces are different; increasing sub-LOÔs  increase resolution, always with a description-dependent consistency between 2N and many-N contributions to observables. 

When treating the ${\Delta}$ isobar as an explicit degree of freedom, our choice had a timely reason and was connected with the issue of resolution in a special way: It was the time of pion factories, programmatically exploring short-range correlations between nucleons. We were back at Brueckner's original problem: What is the short-range part of the NN interaction and how to deal with it? Ambitious as we were, we wanted to create a consistent description for nuclear phenomena at low and intermediate energies, energetically allowing for single-pion ({$\pi$}) production: We described the coupled  NN, {$\pi$}-deuteron (d) and NN{$\pi$} reactions quantitatively \cite{sauer:86}, the {$\Delta$} isobar being dressed to the physical {$\pi$}N resonance. In our approach to nuclear structure, the Fujita-Miyazawa 3N force \cite{fujmiy:57} was built up as an effective one by the succession of two-baryon potentials with channel coupling. In {$\pi$}-nucleus scattering, the dressed {$\Delta$} isobar was considered to move in the medium with an appropriate width for the decay into open {$\pi$}-channels; the required {$\Delta$} optical potential was to be derived microscopically \cite{leeohta:82} in the same way as the N optical potential in the medium. We, and other groups, e.g., Ref. \cite{betzlee:81}, working on NN{$\pi$} dynamics, had partial success, but we never reached a satisfying accuracy for describing observables below and above {$\pi$}-production threshold across the board. This challenging project is now left to the new generation of few-body physicists with new theoretical equipment.

\begin{figure}[!]
\begin{center}
\includegraphics[scale=0.36]{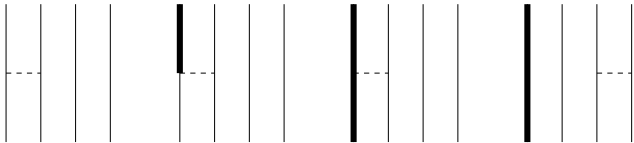}
\end{center}
\caption{\label{fig:Vnd} 
Coupled-channel potential with single {$\Delta$} excitation, shown for the 4N system. The tuned form is CD Bonn + {$\Delta$} \cite {deltuva:03c}. No {$\pi$}-channels and no irreducible many-baryon potentials are introduced; all many-N forces are effective ones arising from iteration of the assumed interaction.}
\end{figure}

We retreated from the physically more realistic version of a coupled-channel potential with active {$\pi$}'s to the simplification without active {$\pi$}'s, shown in Fig.~\ref{fig:Vnd},  presently employed, applicable at low energies and with a competitively precise fit to elastic NN scattering data \cite{deltuva:03c}. We work with consistent effective 2N and many-N forces as shown in Fig.~\ref{fig:mnf} and with  correspondingly consistent currents for electroweak processes. They are derived from all contributing meson exchanges. However, since many-N forces and currents are solely based on {$\Delta$} mediation, they are physically incomplete. 
\begin{figure}[!]
\centering 
\includegraphics[scale=0.48]{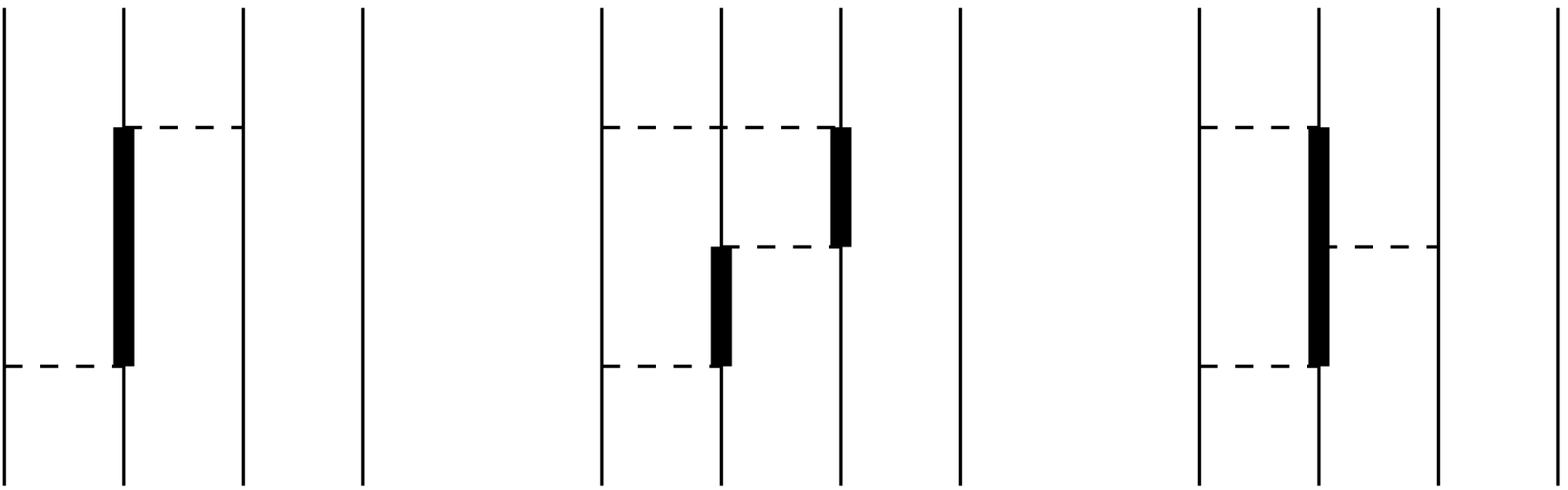}
\end{figure}
\begin{figure}[!]
\centering 
\includegraphics[scale=0.48]{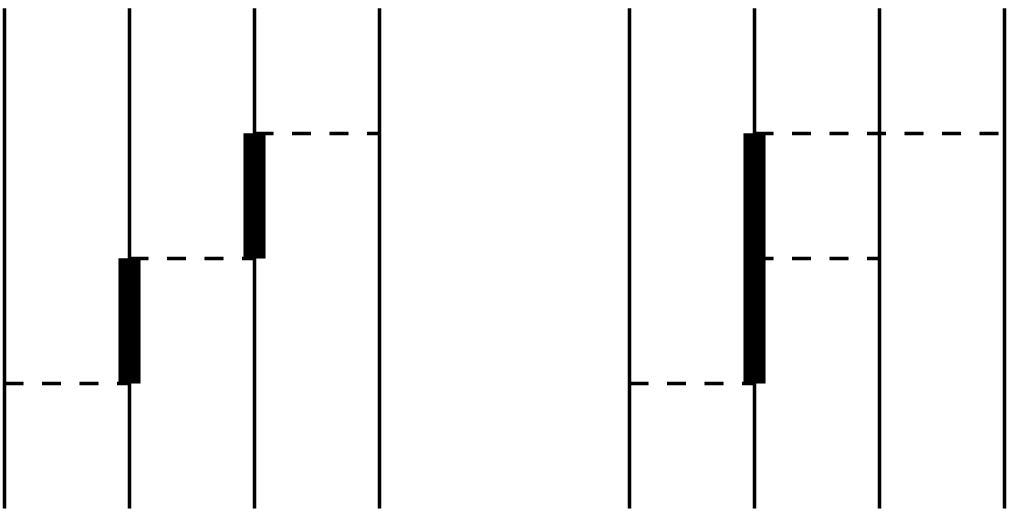}
\caption{{\label{fig:mnf}} Effective 3N and 4N forces. The upper left process yields the Fujita-Miyazawa 3N force \cite{fujmiy:57}.}
\end{figure}

The resulting energies of the 3N and 4N bound states can be split up according to the different dynamic mechanisms; the ${}^4\mathrm{He}$ binding energy confirms the general folklore that the effect of the 4N force, providing a contribution of only 1\% to the total binding, is much smaller in nuclear phenomena at low energies than that of the 3N force, yielding a contribution of about 10\% to the total binding  \cite{deltuva:08a}. The same effects of relative importance reoccur in 3N and 4N scattering. However, we also note: The chosen form of the coupled-channel potential does not have any free parameter which could be adjusted for further tuning. The 3N and 4N binding energies are not fully accounted for.  We do not mind; you may call that an honest approach or a severe limitation for further applications in scattering processes. 


\section{Coulomb in the Description of Scattering \label{sec:3}} 

Most few-nucleon reactions involve charged particles as targets and as beams. Compared to neutron(n)-induced reactions, they are easier to handle experimentally, and they yield more accurate data. However, the long range of the Coulomb potential between protons (p) creates a severe problem for standard scattering theory. 

I revisited this problem only a few years ago \cite{deltuva:08c}; but I did not come from nowhere. For me, the story had already begun at the Delhi Conference, a conference in honor of Mitra's achievements for few-body physics, the only conference which lasted 2 calendar years, 1975 and 1976. I had the assignment for a talk on Coulomb in few-nucleon systems \cite{sauer:76}. My credentials were the problem of subtracting Coulomb from the pp scattering length and from the ${}^3\mathrm{He}$ - ${}^3\mathrm{H}$ mass difference, necessary for determining the amount of charge asymmetry in the nuclear interaction. But the organizers were tough, requiring me also to talk on the inclusion of Coulomb in few-nucleon scattering equations, a subject of which I did not have the slightest experience. I had a full year for preparation, but no internet. One colleague helping out in Germany was Hans H. Hackenbroich, working in configuration-space with the Kohn-variational principle \cite{hacken:70}. But Hackenbroich had to use schematic local forces, and he had a clear conceptual problem: How to go above the break-up threshold? The boundary conditions, e.g., for pd scattering, were not worked out yet in full. Realistic calculations appeared far away. At that time, it was unthinkable that the Pisa group \cite{kievsky:01a} would make this approach to Coulomb so successful, being even able to employ nonlocal potentials \cite{viviani:06}.
  
When discussing Coulomb with him, Hackenbroich also told me about his personal health problem; the advice of his doctor was, not to go to Delhi because of the danger of a liver infection. But he was highly enthusiastic about his work and absolutely confident, that he would be safe with respect to his health by {\it drinking coca cola only}. Nevertheless, he contracted an infection in Delhi and died a few weeks after his return, very young with 43 years of age, a great loss for our field.
\begin{figure}[!]
\begin{center}
\includegraphics[scale=0.35]{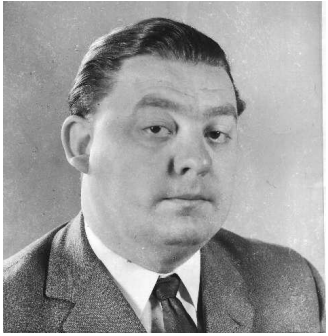}
\end{center}
\caption{\label{fig:Hackenbroich} 
Hans H. Hackenbroich, Universit{\"a}t K{\"o}ln  (1932 - 1976) \cite{hacken:10}.}
\end{figure} 

After Delhi, the interest in screening Coulomb and in using momentum-space integral equations became strong. Screening appears natural, since nature screens it anyhow, and screened Coulomb allows the application of standard scattering theory. The theory problem is that screening does not have the proper unscreened limit, which is, however,  known. Thus, besides screening also a renormalization of the resulting on-shell scattering amplitudes is required. Screening and renormalization has a clean mathematical foundation \cite{taylor:74a}: The convergence of scattering amplitudes  after renormalization is at least in form of a distribution, sufficient for wave packets; furthermore, the replacement of the purely Coulomb part of the on-shell scattering amplitude by the known limit is possible and technically preferable.  

Skipping rigorous mathematics, my simplistic example for explaining the strategy of screening and renormalization is pp scattering, using the two-potential formula; the example will also give me the opportunity to illustrate different strategies possible for tackling the problem practically:  The protons are assumed to interact through the hadronic  $v$ and the Coulomb $w_C$ potentials; Coulomb is screened at a screening radius $R$ to $w_R$. The full pp transition matrix $t$ is then approximated by $t^{(R)}$, i.e., $t^{(R)} = v+w_R + (v+w_R) g_0 t^{(R)}$, $g_0$ being the free resolvent. It is rearranged  to

\begin{equation}  \label{eq:twpot}
t^{(R)} = {t_R}  + {(1+ t_R g_0)} {\tilde{t}^{(R)}} {(1+ g_0 t_R)}
\end{equation}
according to the two-potential formula. $t_R$, i.e., $t_R = w_R + w_R g_0 t_R$, is the transition matrix of screened {\linebreak} Coulomb; the auxiliary operator $ {\tilde{t}^{(R)} = v  + }$ ${v g_{0R} \tilde{t}^{(R)} } $ has as {\it short} a range as the hadronic potential $v$,  $g_{0R}$ being another free resolvent, also including the screened Coulomb potential $w_R$. In  Eq.~(\ref{eq:twpot}) the two terms, not having proper unscreened limits for  $R \to \infty$, are isolated, i.e., the screened Coulomb transition matrix $t_R$ and the Coulomb-distorted  incoming and outgoing wave functions  ${(1+ g_0 t_R)|\mathbf{p} \rangle}$ of momentum $\mathbf{p}$, whose radial dependence is well approximated over the whole screening regime, but which carry the scattering phase of screened Coulomb ${\eta}_{lR}(p)$ and not the one of unscreened Coulomb ${\sigma}_{lR}(p)$. 

Where does the interesting physics occur in this form (\ref{eq:twpot}) of the pp scattering transition matrix? Of course, in the short-range operator, in the 2nd term of the right-hand side, where the crucial interference between hadronic and Coulomb dynamics happens. Preparing the Delhi talk, I had liked Noble's idea of taking Coulomb out explicitly \cite{noble:67} and then working with the full Coulomb resolvent for the short-ranged remainder instead of the corresponding screened one 
$g_{0R}$. I wanted to use Noble's idea for reshuffling equations as in Eq.~(\ref{eq:twpot}), using screening just as a mathematical intermediary, but then calculating the short-range remainder, directly without screening, head on: Working in Coulomb-wave representation appeared the choice, and for that end we studied Coulomb-wave functions in momentum space and their novel singularities  \cite{dreissigacker:79}. 

After Delhi, I gave up the Coulomb problem in few-nucleon reactions: The screening and renormalization hype was swamping the field, and that was not on my agenda yet. In retrospect, the attempt for a head-on calculation of the short-ranged part was my strategic fault at that time; only much later Ref. \cite{ishikawa:09a} made Noble's strategy work in coordinate space.  But the screening and renormalization approach within the AGS quasi-particle expansion \cite{alt:98}, which dominated the Coulomb problem in the framework of momentum-space integral equations for decades, chose such a head-on calculation and thereby became rather intransparent, at least for my taste. In contrast, when now adopting screening \cite{deltuva:08c}, we trust the standard calculational technique of short-ranged potentials for the full amplitude $t^{(R)}$. Subtracting the isolated Coulomb one $t_R$ yields the short-range part of Eq.~(\ref{eq:twpot}) quite easily in the equivalent form $(t^{(R)} -  t_R)$, and renormalization with the appropriate power of the renormalization factor  $z_R(p) = \exp{[-2i({\sigma}_{lR}(p) - {\eta}_{lR}(p))]}$, trivially understood by {\linebreak} straightforward quantum mechanics without any tricky mathematics, is applied to that form of the short-range interference part. Furthermore, our success is also due to a more efficient screening form for Coulomb \cite{deltuva:08c} than used previously.

The procedure for obtaining the pp scattering amplitude by a momentum-space integral equation carries over to the description of 3N and 4N reactions: Coulomb between the protons and with other charged nuclear clusters as d, ${}^3\mathrm{H}$ and ${}^3\mathrm{He}$ is screened, standard scattering theory becomes applicable, the full many-body transition matrix is split into its long-range part and its short-range remainder, and the isolated divergent pieces are either replaced or renormalized. Thus, a controlled mistake is committed for calculational purposes and fully corrected at the end. Convergence is achieved for all considered 3N and 4N reactions at moderate, numerically manageable screening radii $R$ which are, however, large enough to accommodate the hadronic interaction in full; $R$ only gets  prohibitively large when beam energies in elastic scattering and relative final-state (FSI) energies in break-up approach zero. Examples for the convergence are explicitly given in Refs. \cite{deltuva:08c,deltuva:10}. Convergence is internally signalled by the independence of the calculated observables from the screening radius $R$; in pp scattering convergence  is confirmed by the comparison with the exactly known Coulomb-modified phase shifts; in elastic pd scattering convergence is also checked against the corresponding results of the Pisa group in a benchmark comparison  \cite{deltuva:05b}.


\section{Selected Results \label{sec:4}} 

\begin{figure}[!]
\begin{center}
\hspace{0mm} \includegraphics[scale=0.35]{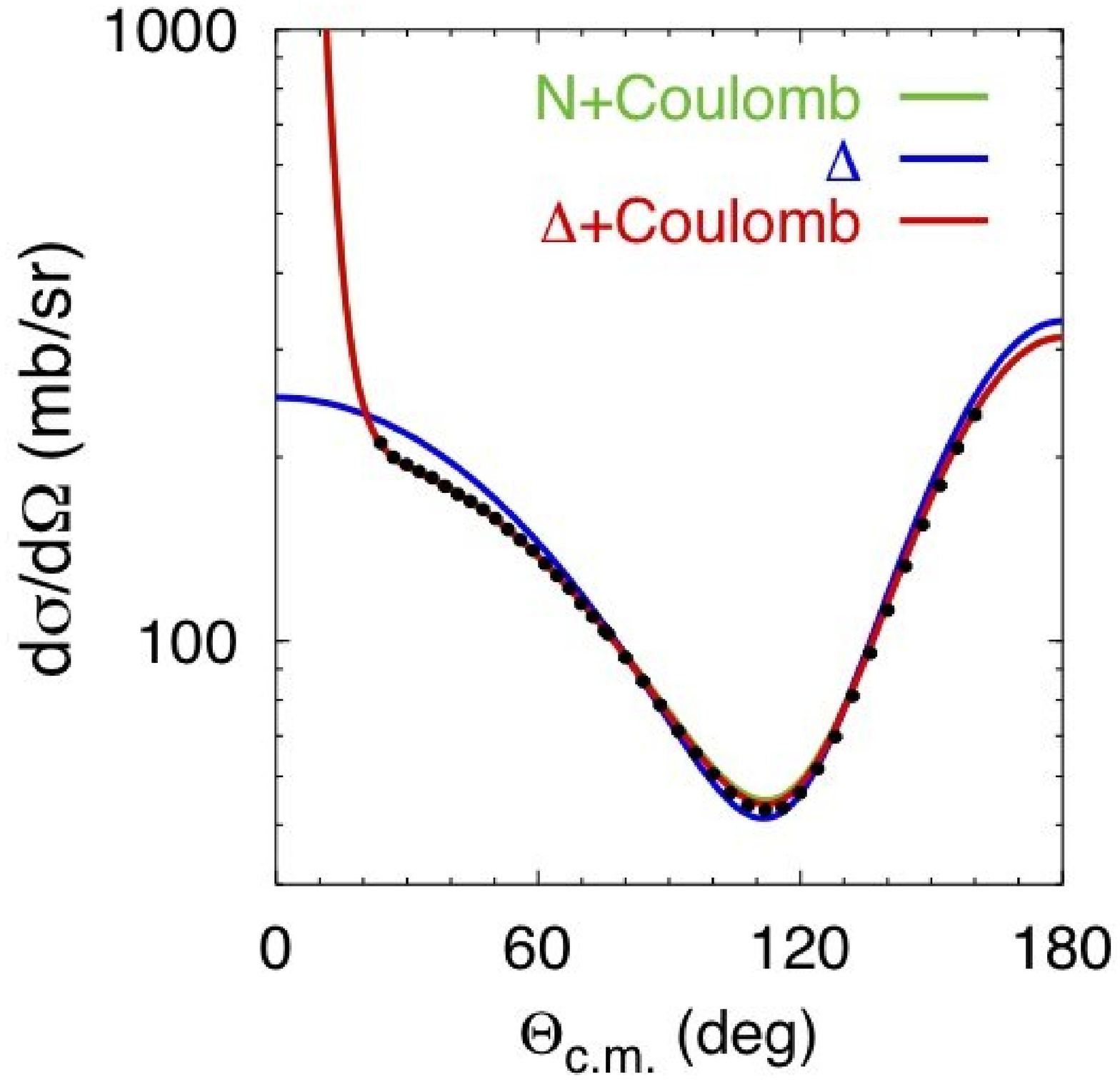} \\  
\vspace{0mm} \hspace{7mm} \includegraphics[scale=0.35]{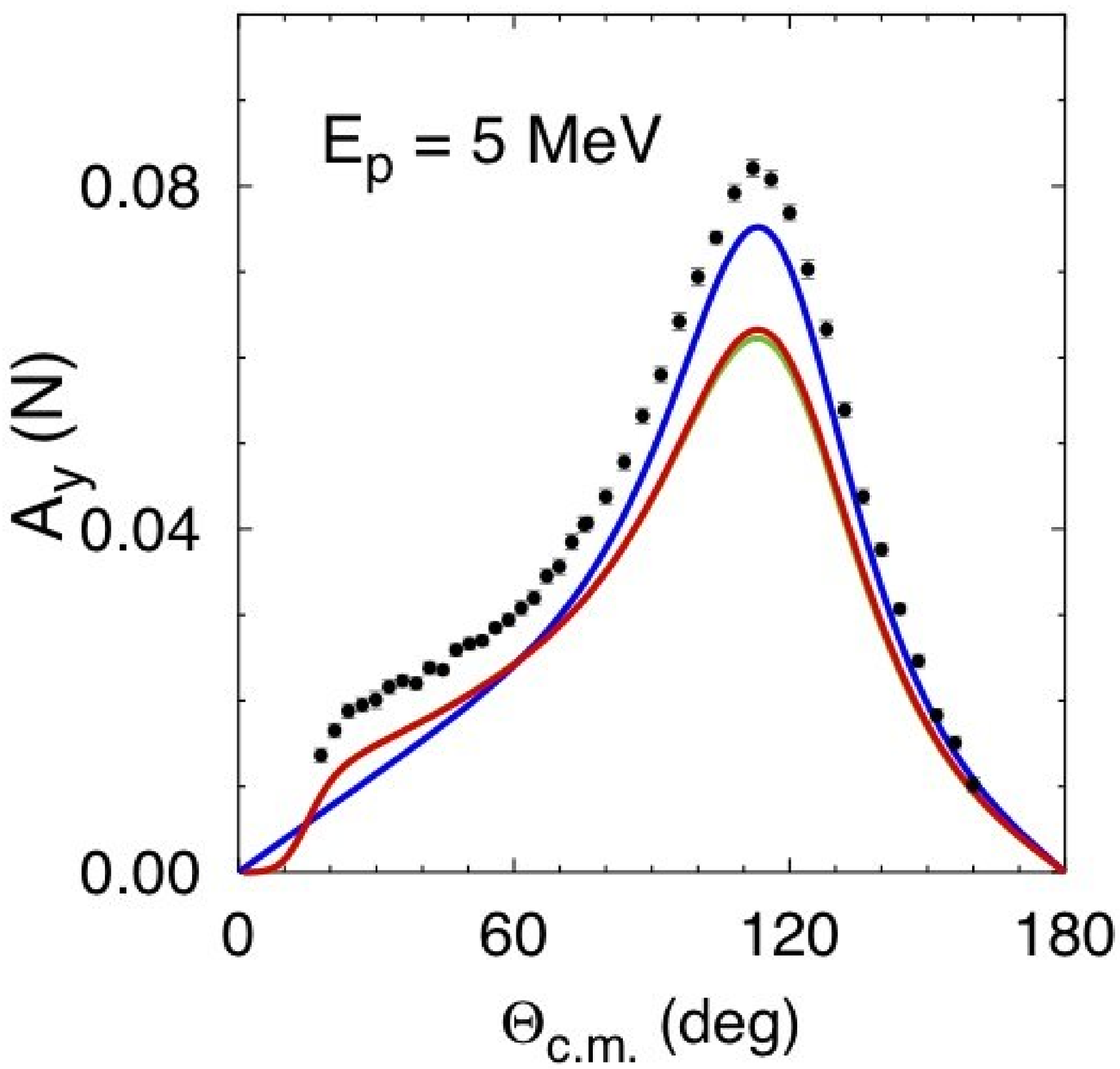}   
\hspace{0mm} \includegraphics[scale=0.60]{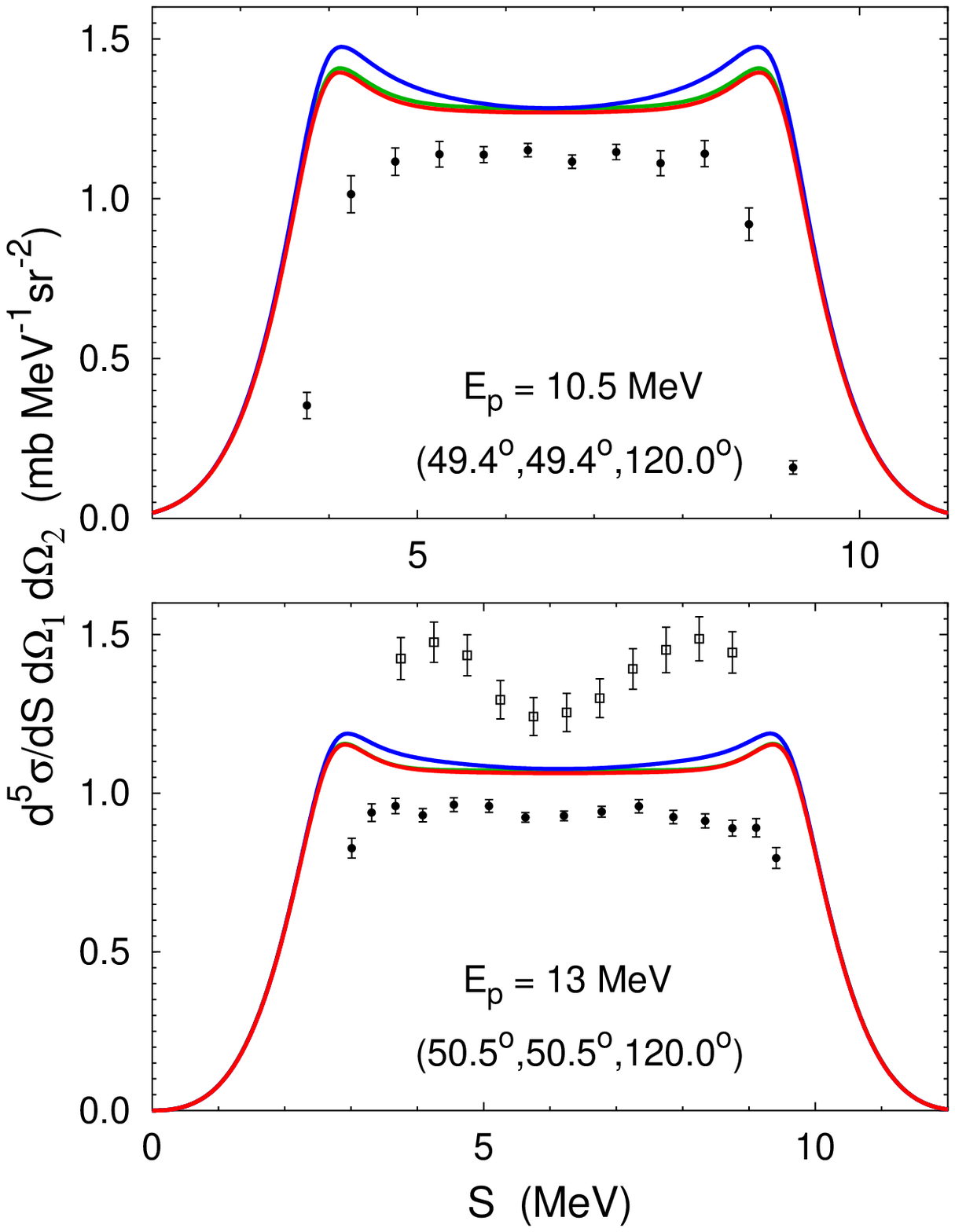}
\end{center}
\caption{{\label{fig:res1}} Elastic pd scattering at $E_p = 5.00$ MeV (top 2 panels) and pd break-up at $E_p = 10.0$ MeV and at $E_p = 13.0$ MeV in space-star kinematics (bottom 2 panels).}
\end{figure}

\begin{figure}[!]
\begin{center}
\includegraphics[scale=0.80]{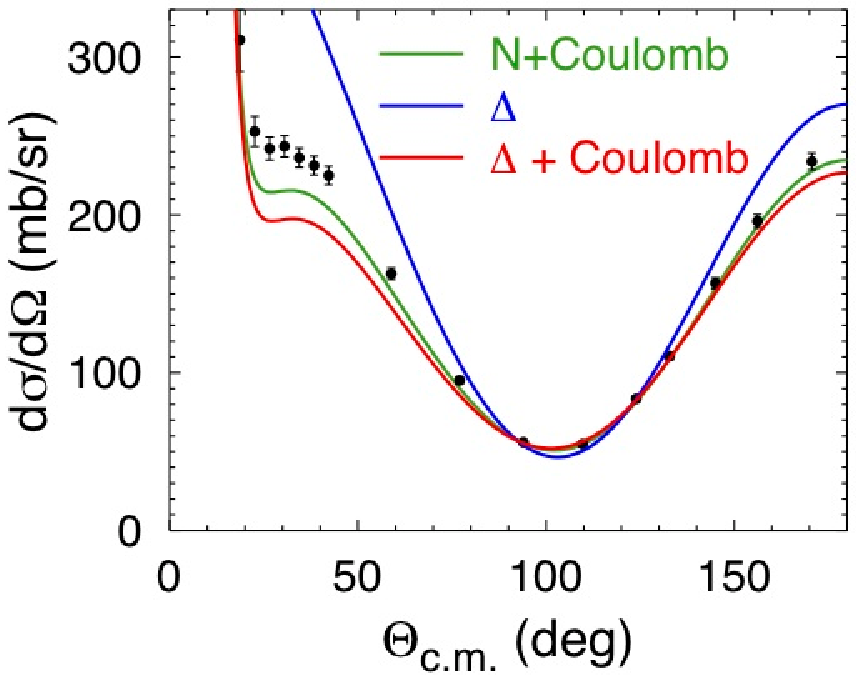} 
\includegraphics[scale=0.80]{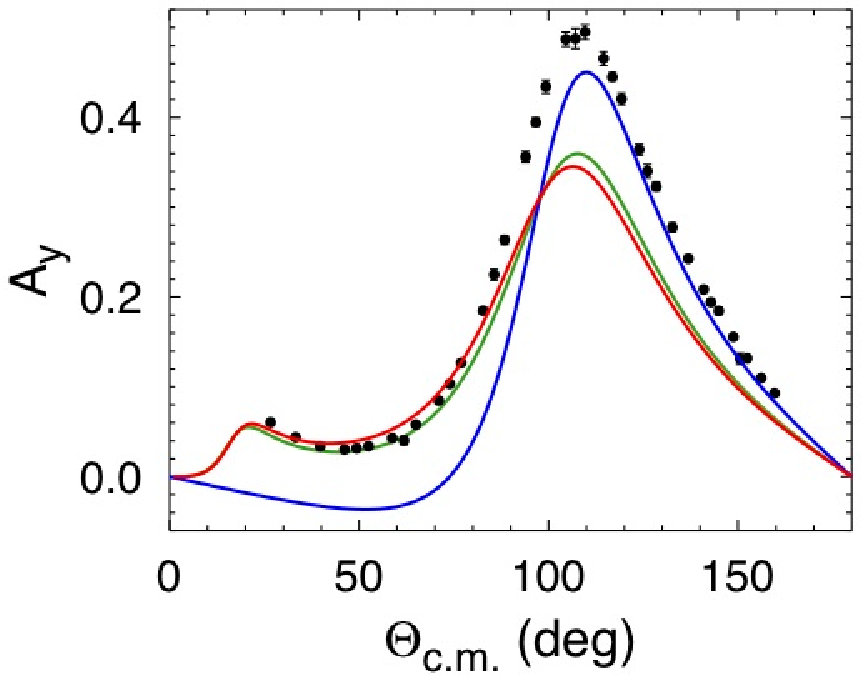} 

\includegraphics[scale=0.53]{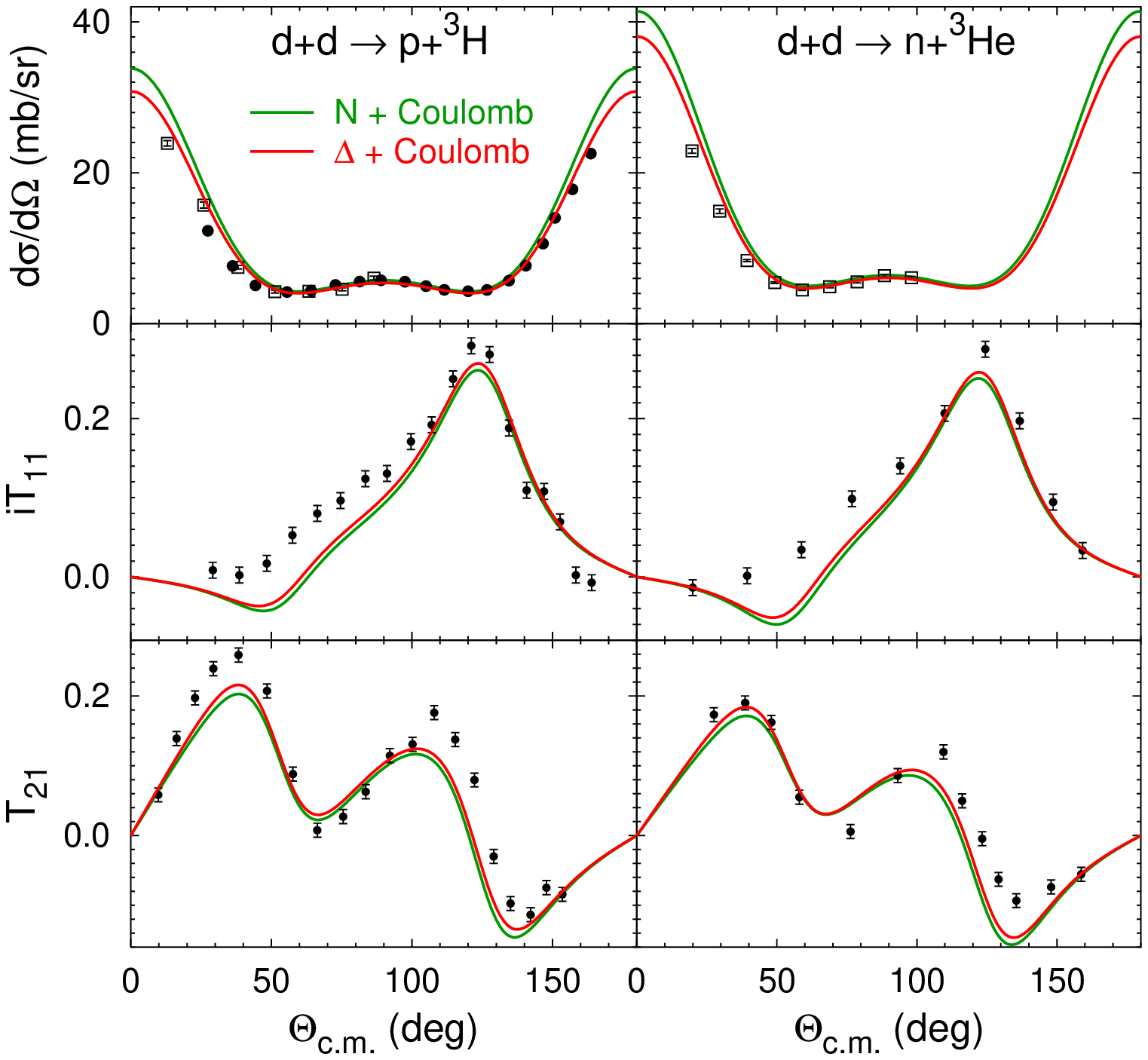}
\end{center}
\caption{{\label{fig:res2}} Elastic p${}^3\mathrm{He}$ scattering at $E_p = 5.54$ MeV (top 2 panels) and the reactions ${\rm dd} \rightarrow {\rm p}{}^3\mathrm{H}$ and ${\rm dd} \rightarrow {\rm n}{}^3\mathrm{He}$ at $E_d = 3.00$ MeV (bottom panels), related by charge symmetry.}
\end{figure}

\begin{figure}[!]
\centering 
\includegraphics[scale=0.60]{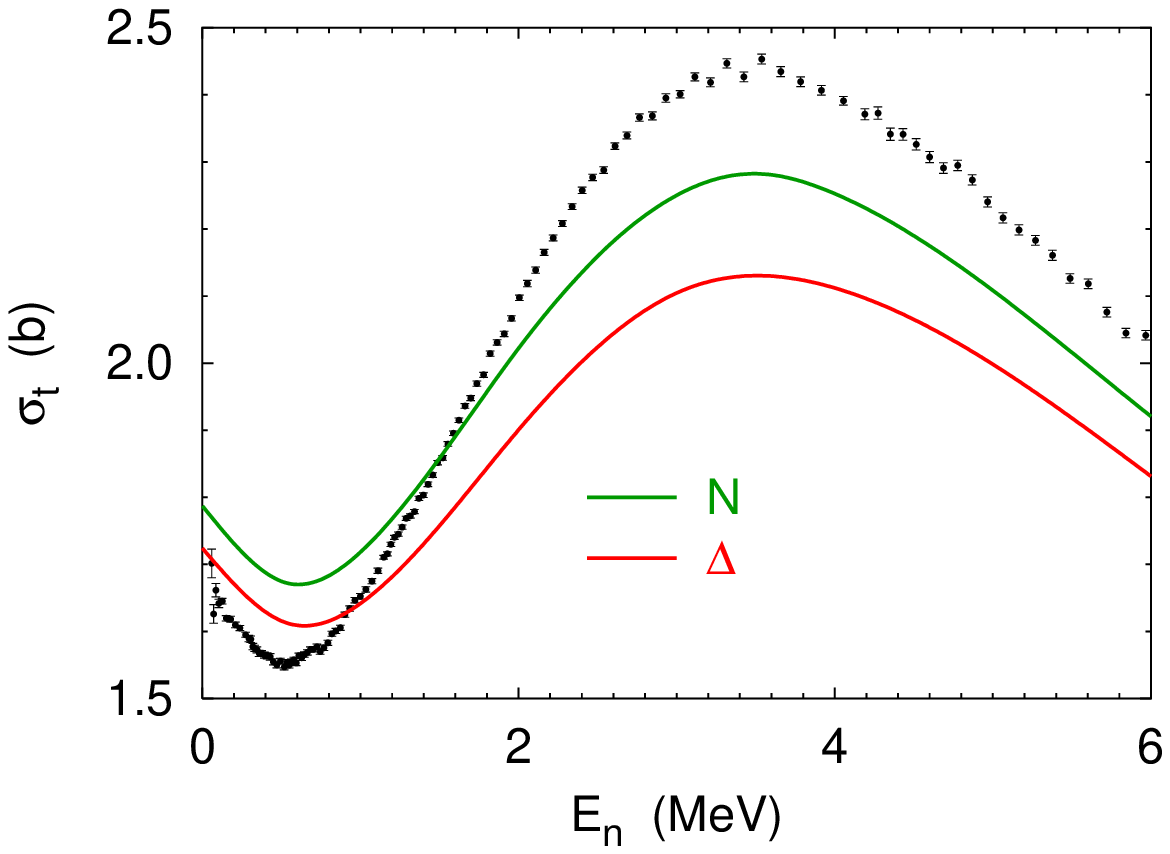} 
\caption{{\label{fig:res3}} Total  elastic n-${}^3\mathrm{H}$ cross section. }
\end{figure}

\begin{figure}[!]
\centering 
\includegraphics[scale=0.50]{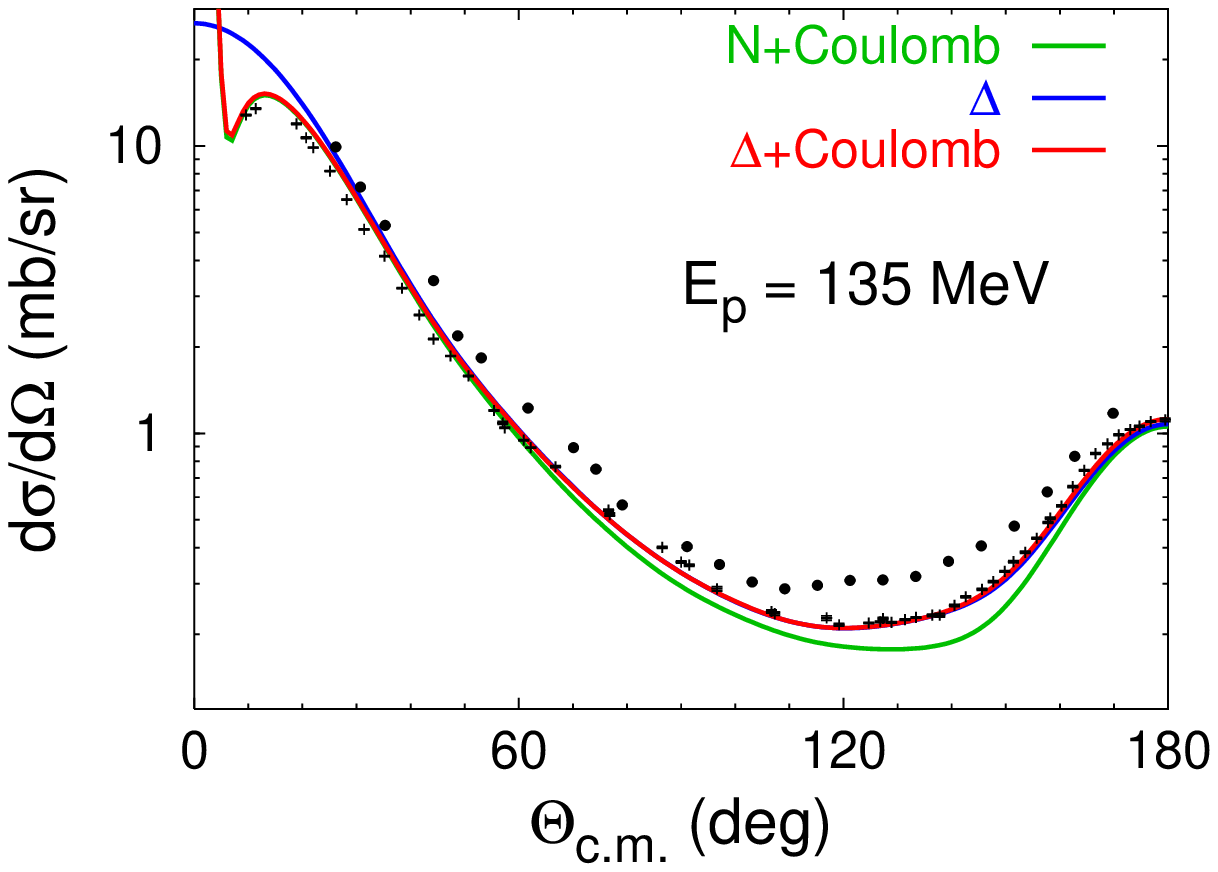} \\
\includegraphics[scale=0.50]{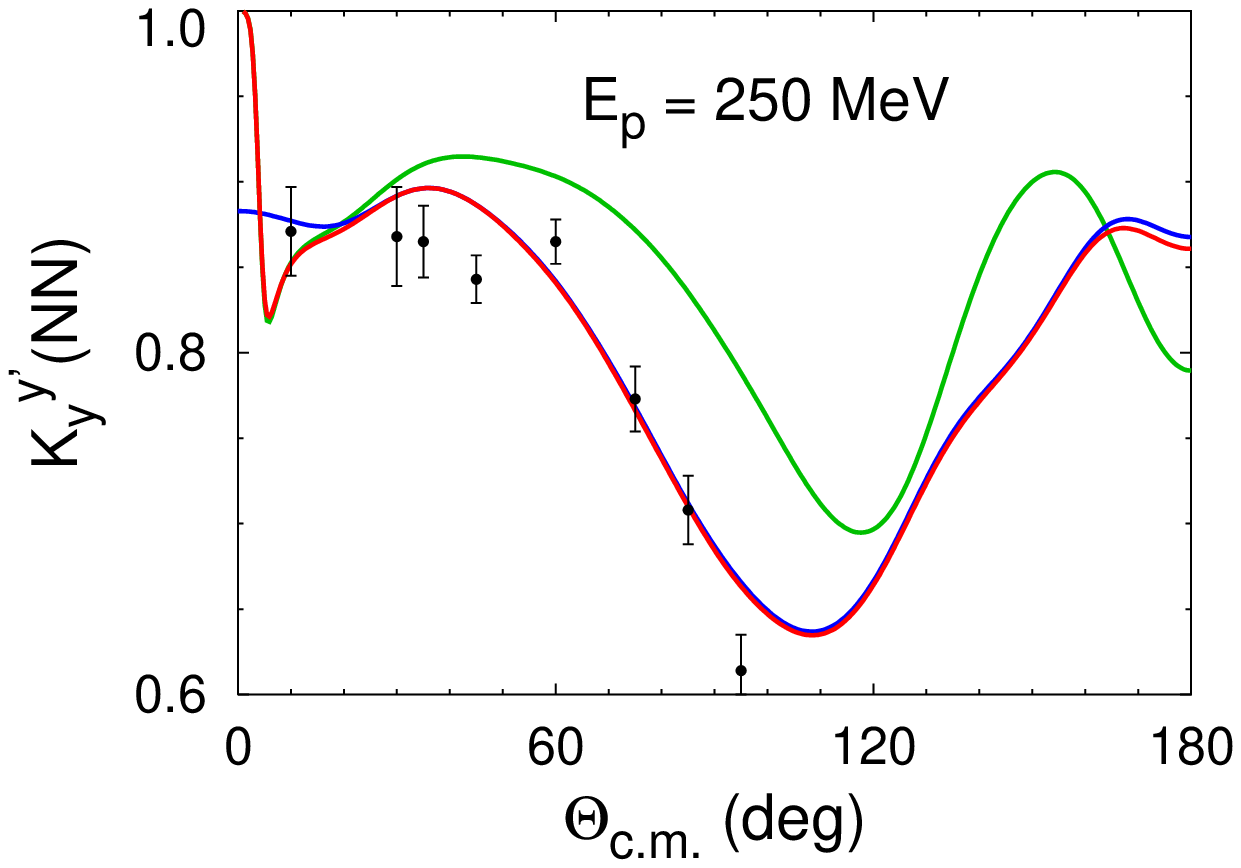} 
\caption{{\label{fig:res4}} Elastic pd scattering at higher energies. }
\end{figure}

\begin{figure}[!]
\centering 
\includegraphics[scale=0.60]{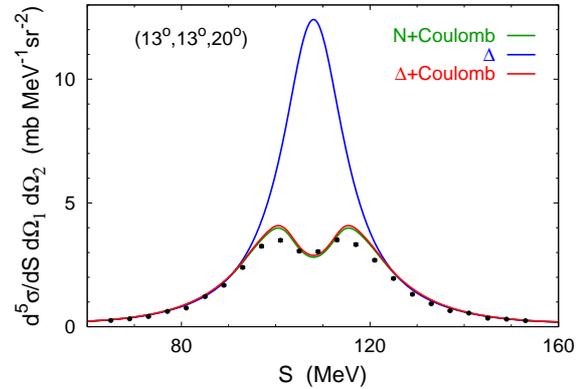} 
\caption{{\label{fig:res5}} Break-up in pd scattering at $E_d = 130$ MeV. }
\end{figure}

Sample results for pd elastic scattering and break-up  \cite{deltuva:05a,kistryn:06} are given in Fig.~\ref{fig:res1}  at low energies and in Figs.~\ref{fig:res4} and \ref{fig:res5} at higher energies. Results for 4N scattering \cite{deltuva:08a} have been obtained so far only for reactions below three-body break-up and are shown in Fig.~\ref{fig:res2} for elastic  
p${}^3\mathrm{He}$ scattering (top) and for the reactions ${\rm dd} \rightarrow {\rm p}{}^3\mathrm{H}$ and ${\rm dd} \rightarrow {\rm n}{}^3\mathrm{He}$ (bottom), related by charge symmetry;  in Fig.~\ref{fig:res3} the total elastic n-${}^3\mathrm{H}$ cross section is shown. The difference between red and blue (green) lines gives the Coulomb (3N-force) effects; in the line-coding $\Delta$ refers to the coupled-channel potential CD Bonn + {$\Delta$}  \cite{deltuva:03c}, N to the purely nucleonic reference potential CD Bonn.


\section{Conclusions and Open Problems \label{sec:5}}

Three theory groups are able to include the Coulomb interaction between protons in their calculation of few-nucleon scattering observables. The Pisa group  \cite{kievsky:01a} {\linebreak} works in configuration space, employing the Kohn-vari\-ati\-onal principle. Ref. \cite{ishikawa:09a} uses coordinate-space integral equations. We made the screening and renormalization approach a reliable technique for momentum-space integral equations \cite{deltuva:08c}.  Each technique has its own advantages. Coulomb effects are important at low energy everywhere according to Figs. \ref{fig:res1} and \ref{fig:res2} top, at all energies in forward elastic scattering according to Fig.~\ref{fig:res4} and in pp-FSI kinematics of 3N break-up according to Fig.~\ref{fig:res5}. A reliable Coulomb treatment is necessary for {\it seeing} the underlying dynamics, but it does not help to understand it. Personally, I am back from where I started from at the time of the Delhi Conference: Then, I had to treat Coulomb reliably, in order to subtract its effect from the pp effective-range parameters and from the ${}^3\mathrm{He}$ - ${}^3\mathrm{H}$ mass difference for a subsequent determination of the amount of charge asymmetry in those observables; it did not help me to understand charge asymmetry better. One could now use the same idea for a comparison of the existing data in pd and nd elastic scattering and break-up and for the 4N reactions ${\rm dd} \rightarrow {\rm p}{}^3\mathrm{H}$ and ${\rm dd} \rightarrow {\rm n}{}^3\mathrm{He}$ of Fig.~\ref{fig:res2} bottom, related by charge symmetry. As determining the amount of charge asymmetry does not help understanding it, in the same way, the proper treatment of Coulomb allows to uncover the nuclear dynamics, but it does not help to understand it.

Conceptually, I emphasized the model-dependence of nuclear potentials. The chosen dynamics yields  $\Delta$-media\-ted effective 2N, 3N and 4N forces, consistent with each other and complete in all meson exchanges. In this description, 4N-force effects are much smaller than 3N-force effects; the 3N force is beneficial for the account of binding energies and of some details of 3N scattering observables at higher energies. Similar results are obtained  by other groups \cite{kievsky:01a,ishikawa:09a,witala:01a} with other chosen dynamics. Many 3N and 4N data are described well by realistic 2N potentials, complemented by a 3N force. In 1972 I had planned 2 to 3 years for my leave of absence from nuclear structure; then all problems of few-nucleon systems should have been resolved, conclusively. But after such a long time, I have to realize that deep physics problems still remain; each solved problem seems to create new ones:

\begin{itemize}

\item 
There are long-standing discrepancies between experiment and theory in 3N and 4N scattering at low energies. One is the $A_y$ puzzle, shown in Figs. ~\ref{fig:res1} and \ref{fig:res2}; the other is the total elastic n-${}^3\mathrm{H}$ cross section, shown in Fig. ~\ref{fig:res3}. However, there is now hope for settling both discrepancies: The Pisa group is able to decrease the discrepancies \cite{viviani:09,viviani:10}, using EFT-based potentials, the Idaho N3LO 2N potential \cite{entem:03} together with the N2LO 3N force \cite{navratil:07}; the forces are not fully consistent yet, but the results are encouraging. Those forces are rather soft ones, with heavily reduced high-momentum components. We have come a long way from the realistic 2N potentials of Brueckner's time with their strongly repulsive core.

\item
The anomaly of pd break-up in space-star kinematics, shown in Fig.~\ref{fig:res1} bottom, remains without any hint of explanation. Though theoretically deter\-mined by quite different parts of the nuclear interaction, the discrepancy shows the same quantitative decrease with energy as the one of  the $A_y$ puzzle. 3N-force effects are small in the present round of calculations, though experiments of this kinematics were originally believed to exhibit large ones. The small effect of Coulomb is confirmed by Ref. \cite{ishikawa:09a}; the large difference between the pd and the nd data at 13 MeV would therefore signal an unbelievably large amount of charge asymmetry. The theoretical overestimate of the pd data is confirmed by Ref.~\cite{glockle:96}, though that calculation is without Coulomb. The EFT-based potentials, which offer hope for the problems with  $A_y$ and with the total elastic n-${}^3\mathrm{H}$ cross section, have not been employed yet for pd break-up.

\item
How can the consistency of data be ensured? There are  two inconsistent data sets for the cross section of elastic pd scattering at $135 ~{\rm MeV}$ p energy. In elastic NN scattering the unitarity of the $S$-matrix is a practical tool for excluding  inconsistent data; no corresponding tool works  yet for 3N scattering above break-up; also this problem is very old. I was post-doc at MIT, when H. Pierre Noyes visited and gave a talk on this subject; his paper \cite{noyes:70} is not of practical value, but the problem he made me aware of remained in my mind as an important one for few-nucleon scattering, and it is still a challenging one, indeed. 
\begin{figure}[!]
\begin{center}
\includegraphics[scale=1.00]{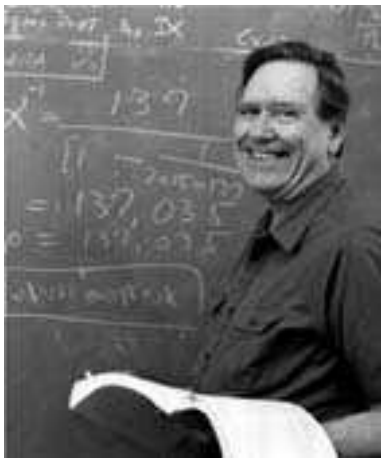}
\end{center}
\caption{\label{fig:Noyes} 
H. Pierre Noyes, SLAC, Stanford University \cite{noyes:10}.}
\end{figure} 

\item 
What is kinematics, what is dynamics in the structure of observables?
\end{itemize}

Convincing answers for the above questions are crucial for the future of the field. \\

Already around 1970 I was told a saying, passed down from our great-great-physics-grandfathers: {\it Never in history before was so much research energy ever spent on one scientific problem as on the two-nucleon interaction.} Over the years, this community has added considerable additional research energy to that topic. What have we learnt about nuclear dynamics from few-nucleon systems?

\begin{itemize}

\item
{\it On the 2N force: } \\
Over the years, I have encountered a number of potentials: Local potentials with singular, finite repulsive and soft cores, momentum-dependent forms, potentials based on one-boson exchange and on 2{$\pi$} exchange, the standard model of nuclear physics and now EFT potentials. At each stage the community got brainwashed to believe that the preached ideology is without alternative and the subject is closed for good. I am pessimistic that we now have reached the ultimate end. And I am disappointed that the systematic inclusion of {$\pi$}-production does not seem to be yet on the agenda for the realistic EFT potentials, so deeply rooted in {$\pi$}-physics.

\item
{\it On the 3N force:} \\
The original claim was: Due to the strong two-nuc\-leon repulsion, three nucleons will never come simultaneously close; thus, the 2{$\pi$} exchange 3N potentials, the companion of the 2{$\pi$} exchange 2N force, is all of the 3N force what nuclear physics will ever need. The community found out painfully that that claim is not true. We introduced and used early {$\Delta$}-mediated many-N forces with shorter ranges. And furthermore, we always wanted to test the 3N force in the 3N system. Now, the community tends to use the 3N observables more and more as data base for fixing the 3N force, postponing its real test to 4N and many-N systems. A long way till now and, clearly, a change of philosophy.

\item
{\it On basic principles of few-nucleon physics:} \\
It is claimed that the EFT philosophy of power-counting may require strategic changes for solving 2N and many-N equations. The summation of all parts of an assumed interaction to all orders by the Schr\"{o}dinger or Faddeev equations may have to be revised. I am deeply worried, and I personally would hesitate to give up those fundamental concepts of successful few-body physics which for example guaranty so holy principles as  the unitarity of the $S$-matrix.  

\end{itemize}


\section{An Afterthought \label{sec:6}} 

This conference in Salamanca is already  the 21st European Few-Body Conference; the first one was 1972 in Budapest. Erich Schmid, T{\"u}bingen, and Ivo Slaus, Zagreb, had the idea of establishing this series of conferences. 

In most evolutionary processes, there is a moment of great danger to the species. That also occurred to the species {\it European Few-Body Physicist}. How did it happen and who saved the species? I think of Erich Schmid and of Konrad Bleuler, now deceased. The occasion was the Delhi Conference. Air fares to Asia were quite expensive in those days. Schmid was running the Secretariat for European Few-Body Physics, and he organized an inexpensive group flight, originating from Frankfurt which participants from Central and Northern Europe joined, quite a sizable group. It was deep winter in the Northern hemisphere, our plane came in from New York, but was held up there for one day by a snow storm. In this snow storm something serious must have happened to the plane; after departure from Frankfurt a technical problem was discovered by Schmid - actually first by his wife -, a bolt fixing a panel on the left wing was gone, the panel moving uncontrolled, a technical problem requiring an immediate stop of the plane in the next available airport, Rome. On the ground and after short inspection, the crew of the plane intended to continue the flight nevertheless without repair. But Schmid who had some hobby interest in and knowledge of the Boeing 747 knew, that a new start with such a defect was considered by Boeing itself as dangerous. He therefore protested to the crew; when unsuccessful, it was Bleuler, who spoke Italian fluently and who communicated to the Italian ground authorities that the request for the start permission was illegal; eventually a successful protest. Thus, the plane got repaired in Rome, and we reached with further delay, but safely Delhi. Accidentally, not so much later, one Boeing 747 of the same airline crashed indeed on the same route due to a technical defect; no one on board survived.
\begin{figure}[!]
\begin{center}
\includegraphics[scale=1.10]{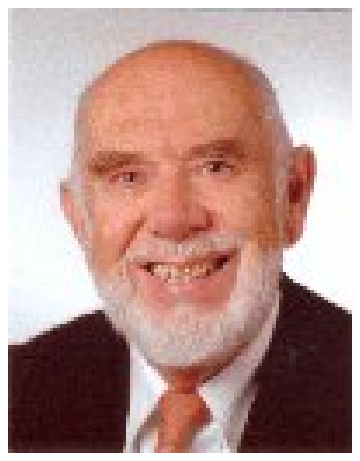}  \hspace{5mm} \includegraphics[scale=0.18]{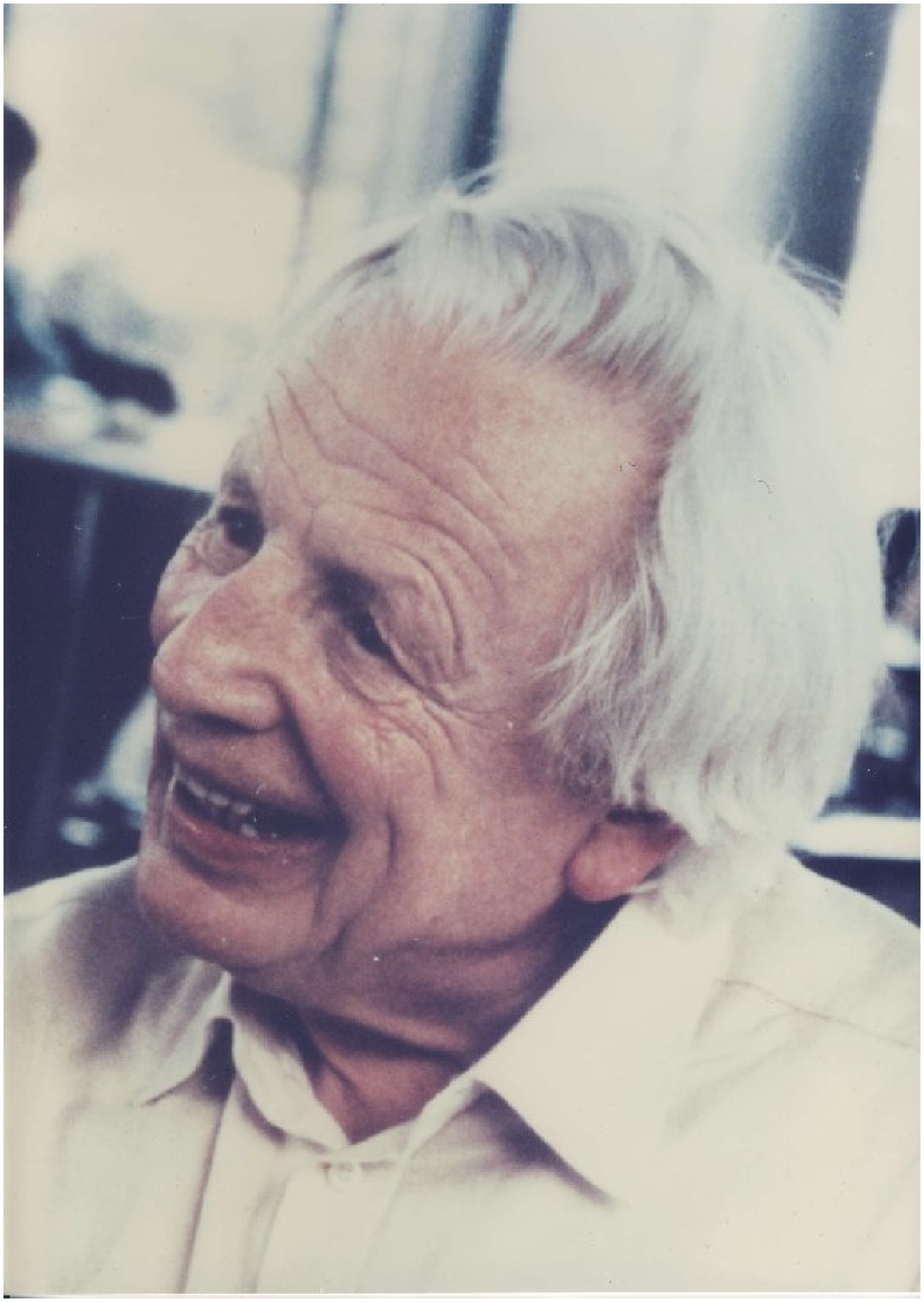}
\end{center}
\caption{\label{fig:SchmidBleuler} 
E. Schmid (left), Universit{\"a}t T{\"u}bingen,  and K. Bleuler (right), Universit{\"a}t Bonn (1912 - 1992) \cite{schmbleu:10}.}
\end{figure} 

The return flight from Delhi provided another surprise for many conference participants. The plane started from Delhi with 5 engines! For people, trained to have a deep respect for symmetries, it was a scary undertaking. Indeed, a Boeing 747 is constructed to carry a 5th engine as load; that special 5th engine had to be brought to the Middle East as replacement for the engine of a stranded plane. Of course, when carrying a 5th engine, the plane should not be so heavily loaded as it was the case on that return flight, perhaps also a borderline risk.
\begin{figure}[!]
\begin{center}
\includegraphics[scale=0.57]{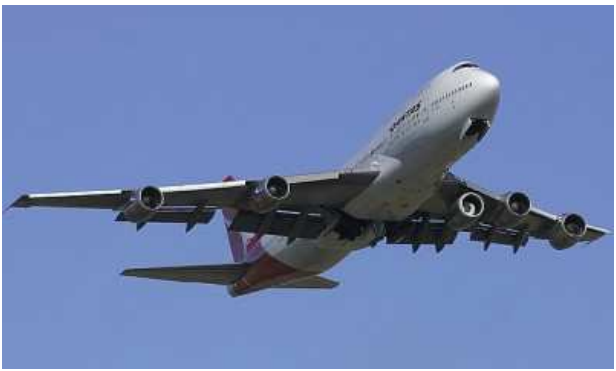}
\end{center}
\caption{\label{fig:5Engine747} 
Example of a Boing 747 with 5 engines \cite{5engine:00}. The airline of the air craft shown has no relation whatsoever to the events told in this talk.}
\end{figure} 

In retrospect, I like to view that incident of the return flight as symbolically providing some extra power to European few-body physics, received at the Delhi Conference. I therefore wish you the same extra power when now returning home from the Salamanca Conference to make few-body physics even more flourishing in future. Have a safe trip home. Being the last speaker, I thank, also in the name of all participants, the organizers of this conference in this beautiful Spanish town and all secretaries behind them for their devoted work. It was a memorable event. Thank you.


\begin{acknowledgements}
The physics results presented in this talk were obtained in a year-long collaboration with A. Deltuva and A.C. Fonseca which the author always enjoyed. The author  thanks A. Kievsky and M. Viviani for enlightening discussions on their recent work. The author is grateful to E. Schmid who refreshed his memory on various details in the history of few-body physics. In the rush of creating this talk and writing it up, the author was unable to ask K. A. Brueckner and H. Pierre Noyes personally for the permission to use their photos; he hopes that they will graciously forgive him. 
\end{acknowledgements}



\end{document}